\documentclass[useAMS,usenatbib]{mn2e}
\usepackage{Abbrev,amsmath,graphicx,xcolor,comment}
\title[Partial Obscuration in AGN]{The Effect of Partial Obscuration on the Luminosity Dependence of 
the Obscured Fraction in Active Galactic Nuclei.}
\author[Jack. H. Mayo \& Andy Lawrence]
       {Jack H. Mayo$^{1}$\thanks{E-mail: jhm@roe.ac.uk} \& Andy Lawrence$^{1}$\\  
       $^1$Institute for Astronomy, SUPA (Scottish Universities Physics Alliance),
         University of Edinburgh, \\ \hspace{1mm}Royal Observatory, Blackford Hill, 
         Edinburgh. EH9 3HJ. \\}

\begin{document}
\maketitle
\begin{abstract} 
Surveys of Active Galactic Nuclei (AGN) in different observational regimes seem to give different answers for the behaviour of the obscured fraction with luminosity. Based on the complex spectra seen in recent studies, we note that partial covering could significantly change the apparent luminosities of many AGN, even after apparent X-ray absorption correction.  We explore whether this effect could reproduce the observed dependence of the obscured fraction on the apparent X-ray luminosities seen between 2--10 keV.  We can reproduce the observed trend in a model where 33 per cent of AGN are unobscured, 30 per cent are heavily buried, and 37 per cent have a range of intermediate partial coverings.   Our model is also tentatively successful at reproducing observed trends in the X-ray vs. infrared luminosity ratio for AGN.
\end{abstract}


\section{Introduction}
Observations in the optical and X-ray regimes of Active Galactic Nuclei (AGN) show them to belong to two distinct groups; the unobscured objects with broad optical emission lines and/or an unattenuated spectrum in the X-ray, and the obscured objects lacking broad optical emission lines and/or with signs of attenuation in the X-ray; the optical and X-ray classifications agree roughly, but not exactly \citep[see][]{Lawrence10}.  When looking at the obscured AGN fraction ($f_{obsc}$), contradictory results have been observed in different regimes; results from X-ray observations (between $2-10$keV) find an $f_{obsc}$ luminosity (L$_{\mathrm{x}}$) dependence \citep{Law82,Ueda03,Hasinger08} though equally other observations contest these findings, \citep[see][]{Dwelly06,Eckart06}. 
Contradictory results are also present in the optical regime; using \texttt{[OIII]} luminosity as a proxy for the intrinsic AGN luminosity, \citet{Simpson05} find a decrease in the number density of Type-II (optically obscured) objects with luminosity, whilst \citet{Lu10} state that this observed anti-correlation can be accounted for purely on the grounds of selection effects and extinction correction.  It is also becoming evident that there is no observed correlation with \texttt{[OIII]} luminosity from sub-mm selected samples (Mayo et al., in prep).

\citet{Lawrence10} have shown that mid Infra-Red (IR) selected samples \citep{deGrijp92,Rush93,Lacy07}, low frequency radio samples \citep{Willott00} and local volume limited samples \citep{Maiolino95} show no such evidence for an $f_{obsc}$ luminosity dependence, with $f_{obsc} \simeq 0.6$ fixed across many decades in luminosity. By going to higher X-ray energies ($>10$keV) one avoids much of the effects of obscuration and so would hope to find a less biased sample of AGN, missing only the most heavily obscured \emph{Compton thick} objects. However, an $f_{obsc}$ vs. L$_\mathrm{X}$ trend is still observed at these energies. Observations in this regime (\emph{Swift}/BAT Survey; \citet{Tueller10, Burlon11}, INTEGRAL mission; \citet{Beckmann09}) include objects that are known to be Compton thick, and \citet{Wang06} argue that by removing such objects the observed $f_{obsc}$ vs. L$_\mathrm{X}$ trend becomes much less significant. As an example, the BAT sample includes \texttt{NGC 1068} listed at its apparent X-ray luminosity, a factor of one hundred less than the known intrinsic X-ray luminosity (log$_{10}$ L$_\mathrm{X}$ = 42.1 vs. 44.6 erg s$^{-1}$ \citep{Pier94}). 

This leads us to question whether complex spectra with partial obscuration, or with complete obscuration in the line of sight accompanied by some kind of scattering back from other sight lines could be a common trait in AGN.

\citet{Turner09} report that for source \texttt{1H 0419-577} - which has a hard X-ray excess compared to extrapolations from models based on previous data - is best fit by a model with partial obscuration by a Compton thick medium with column density log$_{10} $N$_\mathrm{H} \geq 24\,$ cm$^{-2}$ and a covering factor of 70 per cent, and a further intermediate obscurer of column density $22 \leq $log$_{10} $N$_\mathrm{H} \leq 24\,$cm$^{-2}$ and covering factor $\sim16$ per cent. Further, \citet{Winter09} report that half of the objects in the \emph{Swift}/BAT sample require complex spectral fits such as partial covering.  There have been observations made of ``buried objects'' with scattered fractions of less than 1 per cent \citep{Bassani99,Ueda07,Sazonov07}, however it is unknown whether this fraction is scattered or else due to heavy partial obscuration ($\sim99$ per cent covering factor) by a Compton thick medium.  If partial obscuration by a Compton thick medium is commonplace but not always recognised in AGN then the absorption correction will often be under-estimated resulting in wrongly calculated intrinsic X-ray luminosities.

Because most objects have only $2-10$ keV X-ray data, when we assume typical intrinsic partial obscurations in this regime, we begin to see problems: while an unobscured object will appear unobscured to the observer and a heavily (though Compton thin) obscured object will appear obscured and will be corrected accordingly, an observer viewing such an object as \texttt{1H 0419-577} with crude quality data will correct for only the 16 per cent intermediate obscuration, which will lead to an apparent X-ray luminosity only 30 per cent of the intrinsic luminosity, since they will be unaware of the 70 per cent Compton thick partial covering component.  Extrapolating this miscalculation to the ``buried objects'' shows the importance of this problem; when correcting for attenuation in such objects (which realistically will have an intermediate column density obscuring component over the remaining $\sim1$ per cent ``uncovered'' fraction) the resulting inferred intrinsic X-ray luminosity will be two orders of magnitude less than the true intrinsic X-ray luminosity.

We note that many of these problems disappear with data of sufficient quality extended to the hard X-rays, since one can model obscuration effectively in these situations.  For AGN in large surveys however, one normally only has a crude indication to the apparent column and this is where many problems arise.

The aim of this paper is to test whether partial covering effects can reproduce the $f_{obsc}$ vs. $L_\mathrm{X}$ trend seen in $2-10$keV data.  In Section \ref{XLF} we look at how partial covering will effect the X-ray Luminosity Function (XLF) and the apparent  $f_{obsc}$.  In Section \ref{model} we examine simple models with a discrete populations and distributions of covering factors.  In Section  \ref{disc} we discuss our models effect on the IR/X-ray correlation and infer the consequences for the unified scheme.

\section{The Effect of Partial Covering on the  X-ray Luminosity Function}
\label{XLF}
\begin{center}
\begin{figure}
\includegraphics[width=0.45\textwidth]{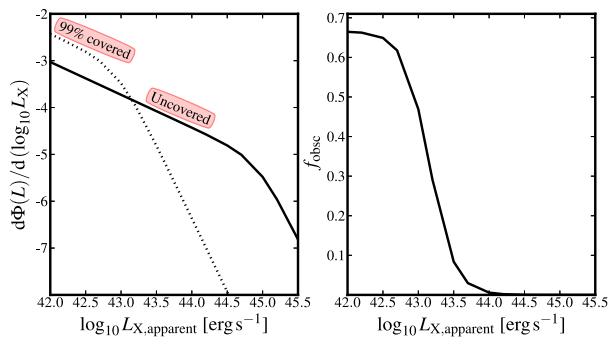}
\caption{The left-hand figure show the intrinsic (uncovered) XLF as taken from \citet{Aird10} (solid line) and the change that is produced in the XLF when every object is covered over 99 per cent of its surface with a Compton thick obscurer (dotted line).  The right hand figure shows the observed $f_{obsc}$ for an intrinsic obscured population making up 50 per cent of all AGN, all being partially covered over 99 per cent of their X-ray surfaces.}
\label{fig:LF}
\end{figure}
\end{center}

By suppressing the X-ray luminosity using a partial obscurer (Compton thick with covering factor $C$) it is possible to shift the luminosity function to be boosted at lower luminosities, and suppressed at high luminosities (since the high luminosity sources are shifted to lower luminosities).  We begin by showing this effect with a simple XLF taken from \citet{Aird10}:
\begin{equation}
\phi(L) = \dfrac{1}{L} \dfrac{d\,\Phi(L)}{d\,\mathrm{log}_{10}(L)} = 
\dfrac{N}{L}\left[ \left(\dfrac{L}{L_*}\right)^{\gamma1} + \left(\dfrac{L}{L_*}\right)^{\gamma2}\right]^{-1}
\end{equation}
Here the best fit parameters they quote are $\mathrm{log}_{10}$($N$)$=-5.1$, $\gamma1=0.7$, $\gamma2=3.14$, $\mathrm{log}_{10}(L_*)=44.96$, characterising the normalisation, low luminosity slope, high luminosity slope and characteristic break luminosity (at the knee of the luminosity function) respectively. For a population with covering factor $C$, the apparent luminosity $K$ will be:
\begin{equation}
K = (1-C)L
\end{equation}
and so if there are N objects intrinsically in luminosity range $dL$, we observe them in range $dK$, where
\begin{equation}
dK = (1-C)dL
\end{equation}
and so the density of objects in observed luminosity space is:
\begin{equation}
\psi(K) = \phi\left(\frac{K}{(1-C)}\right) / (1-C)
\end{equation}
If we consider a simple power law luminosity function $\phi(L) = AL^{-\alpha}$ then
\begin{equation}
\psi(K) = AK^{-\alpha} \cdot (1-C)^{\alpha-1}
\end{equation}
and so for $\alpha < 1$ the effect is to increase the number density, while for $\alpha > 1$ the effect is to decrease the number density. Thus, the adjusted XLF $\psi$, as a function of the \emph{observed} X-ray luminosity $K$, for a fraction $f$ of the total population (i.e. $f < 1$, $\Sigma_i f_i = 1$) is:
\begin{equation}
\psi(K) = \dfrac{N}{K}\dfrac{f}{(1-C)}\left[\left(\dfrac{K}{(1-C)L_*}\right)^{\gamma1}+\left(\dfrac{K}{(1-C)L_*}
\right)^{\gamma2}\right]^{-1}
\end{equation}
which will increase the number density below $L_*$ (where $\alpha \simeq \gamma 1 = 0.7 (< 1.0))$ and decrease the number density above the $L_*$ (where $\alpha \simeq \gamma 2 =3.14 (> 1.0))$.  Consider a simple two population model.  Population-1 is unobscured, with covering factor $c_1=0$, and is a fraction $f_1$ of the total. Population-2 has covering factor $c_2$ and fraction $f_2=(1-f_1)$. Figure \ref{fig:LF} shows the result for $f_1=0.5$ and $c_2=0.99$. The obscured fraction is given by the ratio of the two luminosity functions, and shows a marked trend with apparent L$_\mathrm{X}$. Note that the obscured fraction vs. luminosity shown in the right-hand plot is given by the ratio of the two luminosity functions from the left-hand plot in Figure \ref{fig:LF}. 

\section{Model and Results}
\label{model}
We now discuss a simple model to explain the observed trend in  $f_{obsc}$ vs. $L_\mathrm{X}$. Note that we are not attempting any physical model of the obscuration itself, but simply looking at prescriptions for the distribution of covering factors.  Our model uses populations of partially covered sources to change the apparent X-ray luminosity, with the uncovered fraction for each source allowing through X-ray photons.  The spectra of these sources passing through the ``uncovered'' regions is implicitly taken to be attenuated such that they can be classified as obscured objects, but through a medium sufficiently sparse as to allow through a measurable flux.   While we use the concept of covered fractions and uncovered fractions, it is equally likely that the sources are completely covered along the line of sight, with any X-ray flux coming from a back scattered fraction.  Mathematically these phenomena are identical and as such we use the terms \emph{covering factor} and \emph{covered fraction} to apply to both cases equally.  Note that we are explicitly discussing only the $2-10$k eV energy window, and are ignoring the Compton-hump issues encountered at higher energies.  We explicitly use the term covered fraction here, as opposed to obscured fraction since we want to avoid confusion regarding column densities  and ``amount of obscuration''; the obscuring medium in our partially covered regions is explicitly taken to be Compton thick. We fit our models to data taken from \citet{Hasinger08} and quote best fit parameters in Tables \ref{tab:BF4} and \ref{tab:BF5}.  We perform a Chi-Squared Goodness-of-fit minimisation on all models, with errors quoted as marginalised $1\sigma$ uncertainties.   Throughout we have an uncovered fraction described by $f_1$ and $c_1$ describing the fraction of the total population and the covered fraction of this population, which is $0$ by definition. 

\subsection{Two Population Model}
In Section \ref{XLF} we have shown the effect of a two population model, one partially covered the other uncovered, on the apparent obscured fraction. We now allow the parameters to vary and look for the best fit.  Our free parameters in the model are $f_1$ and $c_{2}$ which are the uncovered fraction and the covered fraction's covering factor respectively (note that $f_{2}$ is constrained as $1-f_1$, and $c_1=0$ by definition). Our best fit parameters are shown in Table \ref{tab:BF4}.  We find this most simple model to be a poor fit the data. We can see from the two population model that a drop in $f_{obsc}$ can be reproduced, but in a much more ``step function'' manner than is observed. Observational data from \citet{Hasinger08} are shown in Figure \ref{fig:Gausspop}

\subsection{Three Population Model}
We increase the complexity of the model by adding another population, this time a population of intermediately covered objects.  Thus we have a model with three populations; population-1 (unobscured, population fraction $f_1$ and partial covering factor $c_0 = 0$), population-2 (lightly covered, population fraction $f_{2}$ and partial covering factor $c_{2}$) and population-3 (heavily covered, population fraction $f_{3}$ and partial covering factor $c_{3}$).  The constraint here is that $f_1+f_{2}+f_{3}$ = 1.0 (so that $f_1$,$f_{2}$ and $f_{3}$ make up the total AGN population).  Our best fit parameters are again shown in Table \ref{tab:BF4}

Despite a $\chi^2_\nu = 2.41$ which is not good enough to accept as a successful model, it is interesting to note two things; firstly that the model predicts an intrinsic $f_{obsc}$ (i.e. $f_2+f_3$) of 55 per cent, similar to the observed optical-IR-radio value, and secondly that a relatively small fraction of intermediately partially covered sources ($f_{3}$, $c_{3}$) is required to produce a measurable difference at log$_{10}\,$L$_\mathrm{X} > 45\,$ erg$\,$s$^{-1}$.

\subsection{Four Population Model}
\label{4pop}
The next logical step is to increase the number of discrete populations to four.  We can see in the \citet{Hasinger08} dataset that there are possibly three plateaux in the data at log$_{10}\,L_\mathrm{X} \sim 43,44.5,46\,$ erg$\,$s$^{-1}$ ; while these may be a feature only present in this dataset, they will undoubtedly be fit by three partially covered populations better than two.  

We fit this six free parameter model and our best fit parameters are given in Table \ref{tab:BF4} where $f_{4}$,$c_{4}$ are the population fraction and partial covering factor for the added \emph{extremely} covered population respectively (that is, a population of sources with in excess of 99 per cent covering factor). 

\begin{center}
\begin{table}
\caption{Best fit parameters with marginalised errors for the discrete population models, along with
reduced chi-squared values $\chi^2_\nu$ and the associated model probabilities, $p$.}
\label{tab:BF4}
\begin{tabular}{|cccc|} \hline 
\textbf{Parameter} & \textbf{2-Pop Model} & \textbf{3-Pop Model} & \textbf{4-Pop Model} \\ 
$c_1$   & $0.000$             & $0.000$              & $0.000              $ \\
$f_1$   & $0.578^{+0.02}_{-0.01}$  & $0.453^{+0.022}_{0.022}$  & $0.313^{+0.0229}_{-0.0114}$ \\
$c_{2}$ & $0.915^{+0.008}_{-0.008}$ & $0.283^{+0.120}_{-0.122}$ & $0.254^{+0.120}_{-0.130}$   \\
$f_{2}$ & $0.422^{+0.01}_{-0.02}$  & $0.126^{+0.010}_{-0.010}$ & $0.068^{+0.0062}_{-0.0061}$  \\
$c_{3}$ &                   -- & $0.959^{+0.004}_{-0.006}$ & $0.941^{+0.008}_{-0.009}$  \\
$f_{3}$ &                   -- & $0.327^{+0.020}_{-0.020}$ & $0.139^{+0.0085}_{-0.0085}$ \\
$c_{4}$ &                   -- & --                   & $0.999^{+0.0002}_{-0.0002}$  \\
$f_{4}$ &                   -- & --                   &   $0.476^{+0.018}_{-0.020}$ \\
$\chi^2_\nu$ & 5.56            & 2.41                 & 1.154                 \\ 
$p$ & $3.1\times10^{-10}$         & 0.005                 & 0.320                \\
\hline
\end{tabular}
\end{table}
\end{center}

We can see with $\chi^2_\nu = 1.15$ that this model fits the data very well.  The model predicts a ``buried fraction'' of $48\pm2$ per cent with 99.9 per cent covering factor, a heavily covered population with $\sim95$ per cent covering factor amounting to 14 per cent of the population, and a smaller population of lightly covered objects making up 7 per cent  of the population with 26 per cent covering factor.  Figure \ref{fig:Gausspop} (left hand plot,
solid line) shows how well this model fits the \citet{Hasinger08} data.

\subsection{Covering Factor Distribution Model}
Phenomenologically, it seems unlikely that a model with discrete populations exhibiting very different covering factors is the correct one; we therefore consider models with a continuous distribution of covering factors. We assume first an uncovered population with $c_1=0$ and fraction $f_1$, and a second population with a range of covering factors such that $f(c)dc$ is the fraction of objects with $c$ in the range $c$ to $c+{\mathrm d}c$. The overall fraction of partially covered objects is $f_2=\int_0^1 f(c) dc$, and the normalisation of $f(c)$ is constrained such that $f_2=1-f_1$. 

(i) {\em Gaussian model.} For the simplest model we took $f(c)$ to be a (half) Gaussian, i.e. with the peak located at $c=1$. It proved impossible to find a satisfactory fit. In order to get the significant population of extremely covered objects (as per $f_4$ in the four-population model) the best fit model is a very narrowly peaked Gaussian, resulting in a step-function-like fit similar to the two population model.

\begin{center}
\begin{figure}
\includegraphics[width=0.45\textwidth]{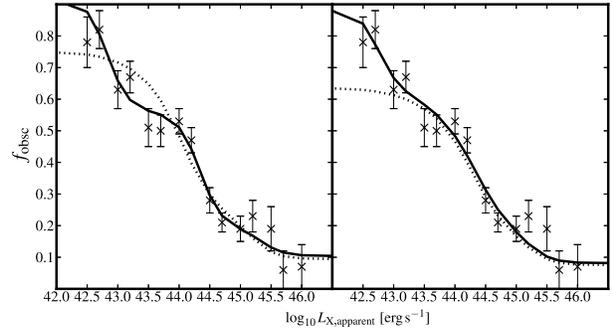}
\caption{Plotting $f_{obsc}$ vs. $L_x$, The left-hand figure shows the best-fit four discrete population model (solid line) and the best-fit Gaussian+constant continuous distribution model (dotted line).  The right-hand figure shows the best-fit log-Gaussian + constant continuous distribution model (solid line) and the best-fit log-Gaussian + constant continuous distribution model excluding the two lowest luminosity bins.  All datapoints shown are taken from the \citet{Hasinger08} X-ray dataset.}
\label{fig:Gausspop}
\end{figure}
\end{center}

(ii) {\em Gaussian plus constant}.  We therefore tried a model in which $f(c) = k + Ae^{-\frac{1}{2}z^2}$ where $z=(c-\mu)/\sigma$. We set $\mu=1$ and constrained the normalisation constant $A$ through the relation $f_2=1 - f_1$; the free parameters are therefore the unobscured fraction $f_1$, constant $k$ and the width of the Gaussian $\sigma$. The best fit parameters are shown in Table \ref{tab:BF5}. The fit is reasonable but not as good as four discrete populations, both of which are shown in the left-hand plot of Figure \ref{fig:Gausspop}

(iii) {\em log Gaussian}. Finally we tried using a Gaussian in $\log_{10} c$, i.e. $f(c)$ as above but with $z=(\log c-\mu_{\log c})/\sigma_{\log c}$. We set $\mu_{\log c}=0$ i.e. $\mu_c=1$, and once again constrained the normalisation so that $f_2=1 - f_1$. We found that to get a good fit to the lowest luminosities we needed to include a third ``buried'' population. We fixed the covering factor of this population at $c_b=0.9989$ and fitted the fraction of this population $f_b$. We therefore have a model with four parameters - the unobscured fraction $f_1$, the buried fraction $f_b$, and for the moderately obscured population, the constant $k$ and the width $\sigma_{\log c}$. The best fit parameters are shown in Table \ref{tab:BF5}, with fit illustrated in Figure \ref{fig:Gausspop}. We note that the requirement for this heavily buried population relies only on the lowest two luminosity points in the \citet{Hasinger08} data. If we ignore these points, we can get a good fit with just the log-Gaussian component (see Figure \ref{fig:Gausspop}).
\begin{center}
\begin{table}
\caption{Best fit parameters with marginalised errors for the Continuous Distribution Models}
\label{tab:BF5}
\begin{tabular}{|ccc|} \hline 
\textbf{Parameter} & \textbf{Gaussian + const.} & \textbf{log-Gaussian + const.}\\ \hline
$f_1$       & $0.381^{+0.01}_{-0.01}$  & $0.333^{+0.007}_{-0.009}$ \\
$\sigma$    & $0.028^{+0.007}_{-0.006}$    & $0.028^{+0.008}_{-0.006}$ \\
$k$           & $0.016^{+0.002}_{-0.002}$    & $0.045^{+0.005}_{-0.01}$ \\
$f_b$       &  --    & $0.300^{+0.02}_{-0.03}$   \\
$c_b$       &  --    & $0.9989$ \\ \hline
$\chi^2_\nu$ & $2.03$  & $1.38$ \\
$p$ & 0.018 & 0.173 \\
\hline
\end{tabular}
\end{table}
\end{center}
\section{Discussion}
\label{disc}
If partial covering explains the apparent  dependence of obscured fraction on X-ray luminosity, then our fits give a reasonably clear picture showing a mixture of clear, buried, and partially covered objects. In the four-population model, 31 per cent of objects are clear and 48 per cent heavily buried; in the intermediate range 7 per cent are 25 per cent covered and 14 per cent are 94 per cent covered.   Likewise, the log-Gaussian model has 33 per cent of objects clear, 30 per cent heavily buried, and 27 per cent with a range of intermediate covering factors. Optically, some of the intermediate objects will probably look like TypeII AGN, and some like Type-I AGN, so it is quite possible that the true obscured fraction is consistent with that seen in optical, MIR, and radio samples (around 60 per cent).  In the rest of this discussion section, we look at the consequences of assuming such a considerable number of partially covered objects.
\subsection{IR/X-ray Ratio}
Ideally one would measure the UV emission to infer the intrinsic luminosity of an AGN.  In reality however, much of the UV emission is attenuated and re-emitted in the IR.  We therefore use the IR as a proxy for the intrinsic luminosity of the AGN, and as such one hopes to see a correlation in the IR vs. attenuation corrected X-ray luminosity ratio.   

\citet{Gandhi09} have shown a tight correlation between X-ray and small-aperture mid-IR luminosity. Their data is reproduced in the upper panel of Figure 3. Note that Gandhi et al made careful efforts to correct for both standard X-ray absorption and Compton thick absorption - partly from the use of hard X-ray data from Suzaku/Integral/SWIFT, and partly from the use of other data such as \texttt{[OIII]} luminosity, which allowed them to estimate the scattered fraction in Compton-thick cases. The degree of correction sometimes necessary is indicated by the two starred points, which show the apparent and corrected luminosities of the archetypal Type 2 AGN, NGC 1068. Overall, the Gandhi et al study shows that when full information and high quality data is available, IR and X-ray emission are tightly correlated.

\begin{center}
\begin{figure}
\includegraphics[width=0.45\textwidth]{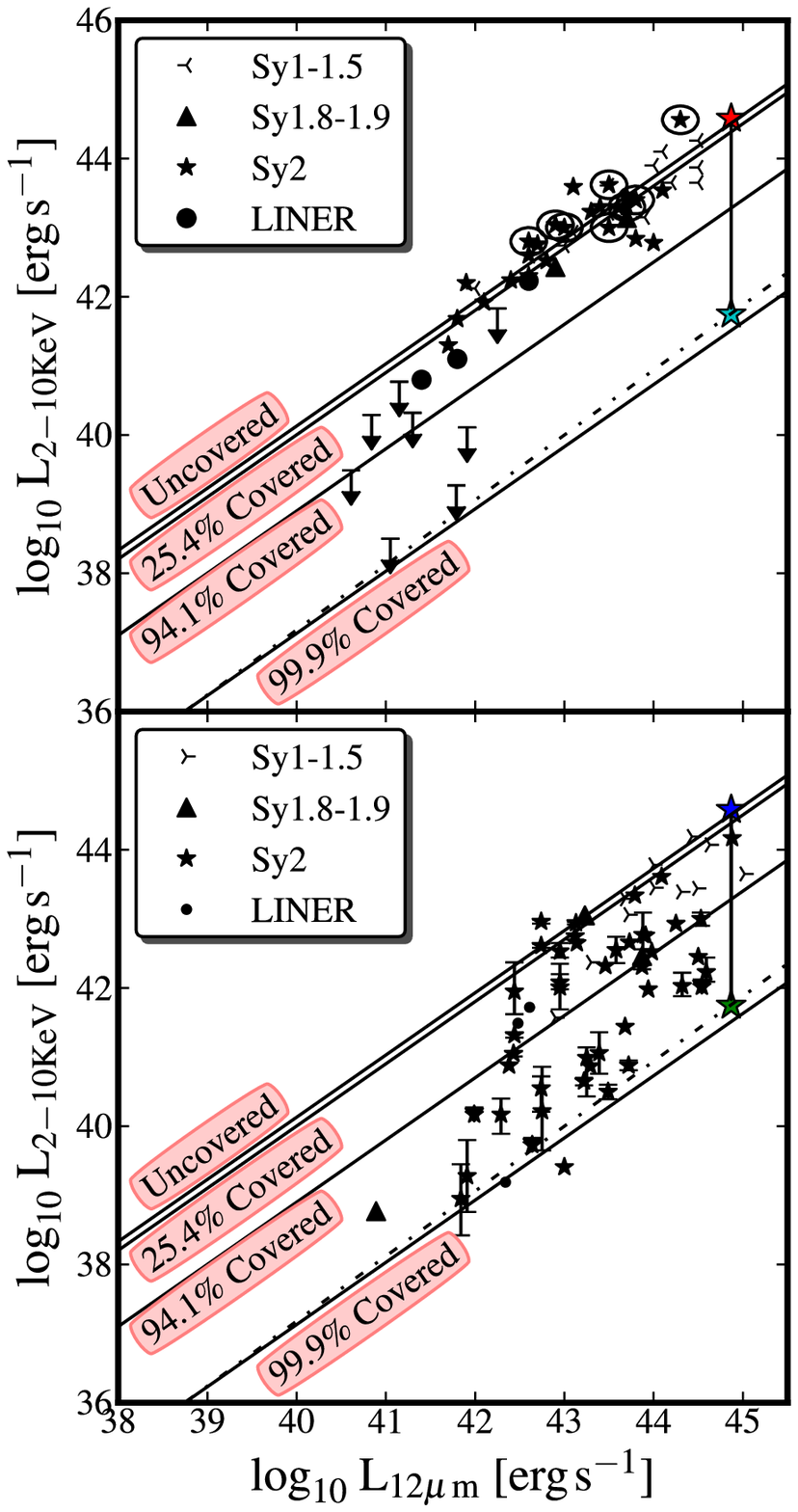}
\caption{A plot of IR vs. ``corrected'' X-ray luminosities for samples of sources taken from \citet{Gandhi09} (top subplot; error bars removed for ease of viewing, sources classified as Compton thick in their work are cicled) and \citet[; no errorbars]{Brightman11b} and \citet[; with error bars]{laMassa11} (bottom subplot) over-plotted with best fit model lines from \citet{Gandhi09} with varying degrees of partial covering.  We have also included \texttt{NGC 1068} (star symbols) at both intrinsic and corrected X-ray luminosities, data being taken from \citet{Pounds06} and \citet{Pier94} for the observed and intrinsic X-ray luminosities respectively, and \citet{Marco03} for the IR luminosity.  The starburst line from \citet{Asmus11} is included as a dot-dash-line to show contamination due to star formation.}
\label{fig:XrayIR}
\end{figure}
\end{center}

A rather different picture is shown in the lower panel of Figure \ref{fig:XrayIR}, which uses data from \citet{laMassa11} and \citet{Brightman11b}. \citeauthor{laMassa11} correct only for the observed X-ray column. \citeauthor{Brightman11b} also use the observed equivalent width of the Fe-K$\alpha$ (6.4keV) line to infer additional Compton thick absorption, using reflection models, but it is far from clear this will find all the additional absorption, and will be much less accurate. These two samples show a large spread in the IR/X-ray ratio. The figure shows the expected effect of various amounts of partial covering, using the Gandhi et al line as the upper envelope. The spread seen is  roughly consistent with the range of partial coverings implied by the model fits of Section \ref{4pop}. It is also clear objects optically classified as Type 1.8-1.9 or Type 2 have a systematically different IR/X-ray ratio. (We note that when there are objects in common between the \citeauthor{laMassa11} and  \citeauthor{Brightman11b} samples, there is soemtimes considerable disagreement on the X-ray luminosity, but perhaps this shows how hard X-ray luminosity correction is).

The contrast between the upper and lower panels of Figure \ref{fig:XrayIR} shows that while it is possible to fully correct X-ray luminosity, standard 2-10keV surveys are unlikely to do so - considerable unrecognised Compton-thick absorption may be present.

As well as the careful correction, the other key difference between the \citet{Gandhi09} study and the  \citet{laMassa11} and \citet{Brightman11b} studies, is that \citet{Gandhi09} used newly measured small aperture mid-IR fluxes, whereas the other two studies use IRAS large aperture fluxes. This could add a spread in the IR/X-ray ratio due to starburst contamination. We have included the X-ray vs IR correlation for starburst galaxies from \citet{Asmus11} in Figure \ref{fig:XrayIR}. \citet{Vasudevan10} have quantified this issue. The correction will vary from object to object of course, but in general no source with log$_{10}$ L$_{\mathrm{X}} > 43$ has greater than 50 percent contamination, while some sources below log$_{10}$ L$_{\mathrm{X}} = 43$ have stellar contamination a factor of several. While this is therefore important, it does not explain the effect shown.

\subsubsection{Star Formation}
Could the IR/X-ray luminosity ratio observed be accounted for by a difference in star formation rates? Evidence for Type-II AGN having higher star formation rates than their Type-I counterparts is disputed in the literature; results from \citet{Maiolino95,Hilner09} claim a statistically significant difference (by up to a factor of 50 percent), while results from \citet{Netzer09,Melendez08}, Mayo et al., (in prep) show no such difference in star formation rates. If such a discrepancy does exist between star formation rates then this would not be seen in the IR luminosities of the \citet{Gandhi09} data in Figure \ref{fig:XrayIR}, since the data here are from small apertures probing only the central regions.  Even if we allow for this effect, a 50 percent difference in star formation rates between Type-I and Type-II objects does not account for the 2-3 order of magnitude spread in IR luminosity in the \citet{Brightman11b} and \citet{laMassa11} samples.

\subsubsection{Variability}
Could variability account for the spread in X-ray vs. IR luminosities that are observed in the \citet{Brightman11b} and \citet{laMassa11} datasets?  Infrared variability is unimportant.  Although near-IR variability in Seyfert galaxies is well known \citep[e.g.][]{Suganuma06} it is typically only a few tenths of a magnitude;  in the mid-IR, variations are seen on a timescale of years with Spitzer, but they are of typical size 0.1 mag \citep{Kozlowski10}.  In the X-ray regime variability is a common trait, with flux variations of order 50 percent on intra-day timescales \citep[e.g.][]{Uttley03,Arevalo08,Breedt09}.  The absolute variability measured in a number of objects cannot feasibly account for the 2-3 order of magnitude spread spread seen in the data.  However, variability is an issue that will hinder measurements of X-ray vs. IR luminosities which can only be overcome by simulataneous measurements in both regimes. 

\subsection{The Compton Thick Fraction, the ``Buried'' Fraction and the X-ray Background}
Observations with \emph{Swift}/BAT \citep{Tueller10} and INTEGRAL \citep{Krivinos05} resolve around $1-2$ per cent of the XRB at $>10$keV.  XRB constrained population synthesis models require a Compton thick fraction in the range $15-25$ per cent \citep[e.g.][]{Beckmann09,Tueller11,Akylas09,Brightman11a}, though increasingly models are favouring the higher end of this range \citep{Gilli07,Akylas12}.  NuSTAR is capable of resolving between $30$ and $50$ per cent  \citep{Von07} and will bring renewed constraints on the Compton thick fraction.

Our model implies a ``buried'' fraction of $\sim $30 per cent. If we identify these as traditional Compton thick objects, then this number is roughly consistent with XRB models, although somewhat larger than these models find. However, our estimate of the number of such buried objects is sensitive to the lowest two points in the data from \citet{Hasinger08} that we have been using.  

\subsection{The Unified Scheme}
It is important to know whether $f_{obsc}$ really does depend on luminosity or not; some models do predict such a trend (e.g. the receding torus model \citep{Lawrence91, Simpson05}), whilst others do not (e.g. misaligned disc model \citep{Lawrence10}).  We have shown that a trend in the \emph{observed} $f_{obsc}$ vs. L$_\mathrm{X}$ can be replicated, implying that it is not necessary to have a luminosity dependent \emph{intrinsic} fraction of uncovered sources.  In a simple torus model, the fraction of objects with at least some degree of obscuration ($\sim$57 per cent) corresponds to a torus opening half-angle of $55^\circ$. On the other hand the number of buried objects ($\sim$ 30 per cent)  corresponds to a opening half-angle of $72^\circ$. In between these two angles would be a rather substantial graded region where the torus is not opaque, but produces partial covering. This is quite hard to understand in a traditional smooth torus model, but may happen fairly naturally in a clumpy torus model \citep[e.g.][and references therein]{Stalevski12}. If the mean number of clouds in the line of sight is $\mu$, then assuming that along different sight lines this quantity is Poisson distributed, then the fraction of sight lines with one or more clumps, i.e. the covering factor, will be $c=1-e^{-\mu}$.  So for example $\mu=1$ would give $c=0.63$ whereas $\mu=7$ would give $c=0.999$.  This general point about clumpiness also applies to alternatives such as disc wind or misaligned disc models. 

\subsection{Future Work}
In order to be considered viable, this model must be extended to include optical obscuration effects, allow for the soft X-ray excess and day-to-day variability seen in many sources.  The model could then be either evolved using XLF evolution models to see whether the relative fractions of each population remain constant over time, or else infer constraints on the unobscured XLF at high redshifts.  Perhaps more importantly, we need to develop a physical model which predicts what the distribution of covering factors might be.  These aims are far outwith the scope of this paper, which simply address the obscured fraction X-ray luminosity dependence in light of recent spectral observations of individual AGN in the $2-10$keV X-ray regime.

\section{Conclusions}
The obscured fraction of AGN is seen to decrease as a function of luminosity in the X-ray regime; this is at odds with volume limited and IR samples.  We have analyzed a model whereby partial covering of the X-ray source in AGN produces a shift in the population density in the observed X-ray luminosity function, and with it a decline in the observed number of obscured AGN at high luminosities, replicating the trend seen in the X-ray.  Our model implies a fixed intrinsic uncovered fraction of 33 per cent and obviates the need for luminosity dependence of key parameters in the Unified Model of AGN.  

Furthermore, we find a ``buried'' fraction, with 99.9 per cent covering factor representing 30 per cent  of the population.   The remaining fraction, making up 37 per cent of the population, are subject to a distribution of covering factors, reproducing the steady decline of obscured AGN with luminosity.  Our model agrees well with IR/X-ray data implying that the intrinsic luminosity of a significant fraction of AGN is consistently underestimated.

\section*{Acknowledgements}
JHM acknowledges the support of the Science and Technology Facilities Council (STFC) via the award of an STFC Studentship.

\bibliographystyle{mnras}
\bibliography{bibliography}

\begin{thebibliography}{48}
\expandafter\ifx\csname natexlab\endcsname\relax\def\natexlab#1{#1}\fi

\bibitem[{Aird} et~al.(2010){Aird}, {Nandra}, {Laird} et~al.]{Aird10}
{Aird} J., {Nandra} K., {Laird} E.~S., et~al., 2010, \mnras, 401, 2531

\bibitem[{Akylas} et~al.(2012){Akylas}, {Georgakakis}, {Georgantopoulos},
  {Brightman} \& {Nandra}]{Akylas12}
{Akylas} A., {Georgakakis} A., {Georgantopoulos} I., {Brightman} M., {Nandra}
  K., 2012, \aap, 546, A98

\bibitem[{Akylas} \& {Georgantopoulos}(2009)]{Akylas09}
{Akylas} A., {Georgantopoulos} I., 2009, \aap, 500, 999

\bibitem[{Ar{\'e}valo} et~al.(2008){Ar{\'e}valo}, {Uttley}, {Kaspi}, {Breedt},
  {Lira} \& {McHardy}]{Arevalo08}
{Ar{\'e}valo} P., {Uttley} P., {Kaspi} S., {Breedt} E., {Lira} P., {McHardy}
  I.~M., 2008, \mnras, 389, 1479

\bibitem[{Asmus} et~al.(2011){Asmus}, {Gandhi}, {Smette}, {H{\"o}nig} \&
  {Duschl}]{Asmus11}
{Asmus} D., {Gandhi} P., {Smette} A., {H{\"o}nig} S.~F., {Duschl} W.~J., 2011,
  \aap, 536, A36

\bibitem[{Bassani} et~al.(1999){Bassani}, {Dadina}, {Maiolino}
  et~al.]{Bassani99}
{Bassani} L., {Dadina} M., {Maiolino} R., et~al., 1999, \apjs, 121, 473

\bibitem[{Beckmann} et~al.(2009){Beckmann}, {Soldi}, {Ricci}
  et~al.]{Beckmann09}
{Beckmann} V., {Soldi} S., {Ricci} C., et~al., 2009, \aap, 505, 417

\bibitem[{Breedt} et~al.(2009){Breedt}, {Ar{\'e}valo}, {McHardy}
  et~al.]{Breedt09}
{Breedt} E., {Ar{\'e}valo} P., {McHardy} I.~M., et~al., 2009, \mnras, 394, 427

\bibitem[{Brightman} \& {Nandra}(2011{\natexlab{a}})]{Brightman11a}
{Brightman} M., {Nandra} K., 2011{\natexlab{a}}, \mnras, 413, 1206

\bibitem[{Brightman} \& {Nandra}(2011{\natexlab{b}})]{Brightman11b}
{Brightman} M., {Nandra} K., 2011{\natexlab{b}}, \mnras, 414, 3084

\bibitem[{Burlon} et~al.(2011){Burlon}, {Ajello}, {Greiner}, {Comastri},
  {Merloni} \& {Gehrels}]{Burlon11}
{Burlon} D., {Ajello} M., {Greiner} J., {Comastri} A., {Merloni} A., {Gehrels}
  N., 2011, \apj, 728, 58

\bibitem[{de Grijp} et~al.(1992){de Grijp}, {Keel}, {Miley}, {Goudfrooij} \&
  {Lub}]{deGrijp92}
{de Grijp} M.~H.~K., {Keel} W.~C., {Miley} G.~K., {Goudfrooij} P., {Lub} J.,
  1992, \aaps, 96, 389

\bibitem[{Dwelly} \& {Page}(2006)]{Dwelly06}
{Dwelly} T., {Page} M.~J., 2006, \mnras, 372, 1755

\bibitem[{Eckart} et~al.(2006){Eckart}, {Stern}, {Helfand}, {Harrison}, {Mao}
  \& {Yost}]{Eckart06}
{Eckart} M.~E., {Stern} D., {Helfand} D.~J., {Harrison} F.~A., {Mao} P.~H.,
  {Yost} S.~A., 2006, \apjs, 165, 19

\bibitem[{Gandhi} et~al.(2009){Gandhi}, {Horst}, {Smette} et~al.]{Gandhi09}
{Gandhi} P., {Horst} H., {Smette} A., et~al., 2009, \aap, 502, 457

\bibitem[{Gilli} et~al.(2007){Gilli}, {Comastri} \& {Hasinger}]{Gilli07}
{Gilli} R., {Comastri} A., {Hasinger} G., 2007, \aap, 463, 79

\bibitem[{Hasinger}(2008)]{Hasinger08}
{Hasinger} G., 2008, \aap, 490, 905

\bibitem[{Hiner} et~al.(2009){Hiner}, {Canalizo}, {Lacy} et~al.]{Hilner09}
{Hiner} K.~D., {Canalizo} G., {Lacy} M., et~al., 2009, \apj, 706, 508

\bibitem[{Koz{\l}owski} et~al.(2010){Koz{\l}owski}, {Kochanek}, {Stern}
  et~al.]{Kozlowski10}
{Koz{\l}owski} S., {Kochanek} C.~S., {Stern} D., et~al., 2010, \apj, 716, 530

\bibitem[{Krivonos} et~al.(2005){Krivonos}, {Vikhlinin}, {Churazov},
  {Lutovinov}, {Molkov} \& {Sunyaev}]{Krivinos05}
{Krivonos} R., {Vikhlinin} A., {Churazov} E., {Lutovinov} A., {Molkov} S.,
  {Sunyaev} R., 2005, \apj, 625, 89

\bibitem[{Lacy} et~al.(2007){Lacy}, {Petric}, {Sajina} et~al.]{Lacy07}
{Lacy} M., {Petric} A.~O., {Sajina} A., et~al., 2007, \aj, 133, 186

\bibitem[{LaMassa} et~al.(2011){LaMassa}, {Heckman}, {Ptak} et~al.]{laMassa11}
{LaMassa} S.~M., {Heckman} T.~M., {Ptak} A., et~al., 2011, \apj, 729, 52

\bibitem[{Lawrence}(1991)]{Lawrence91}
{Lawrence} A., 1991, \mnras, 252, 586

\bibitem[{Lawrence} \& {Elvis}(1982)]{Law82}
{Lawrence} A., {Elvis} M., 1982, \apj, 256, 410

\bibitem[{Lawrence} \& {Elvis}(2010)]{Lawrence10}
{Lawrence} A., {Elvis} M., 2010, \apj, 714, 561

\bibitem[{Lu} et~al.(2010){Lu}, {Wang}, {Zhou} \& {Wu}]{Lu10}
{Lu} Y., {Wang} T., {Zhou} H., {Wu} J., 2010, ArXiv e-prints

\bibitem[{Maiolino} \& {Rieke}(1995)]{Maiolino95}
{Maiolino} R., {Rieke} G.~H., 1995, \apj, 454, 95

\bibitem[{Marco} \& {Brooks}(2003)]{Marco03}
{Marco} O., {Brooks} K.~J., 2003, \aap, 398, 101

\bibitem[{Mel{\'e}ndez} et~al.(2008){Mel{\'e}ndez}, {Kraemer}, {Schmitt}
  et~al.]{Melendez08}
{Mel{\'e}ndez} M., {Kraemer} S.~B., {Schmitt} H.~R., et~al., 2008, \apj, 689,
  95

\bibitem[{Netzer}(2009)]{Netzer09}
{Netzer} H., 2009, \mnras, 399, 1907

\bibitem[{Pier} et~al.(1994){Pier}, {Antonucci}, {Hurt}, {Kriss} \&
  {Krolik}]{Pier94}
{Pier} E.~A., {Antonucci} R., {Hurt} T., {Kriss} G., {Krolik} J., 1994, \apj,
  428, 124

\bibitem[{Pounds} \& {Vaughan}(2006)]{Pounds06}
{Pounds} K., {Vaughan} S., 2006, \mnras, 368, 707

\bibitem[{Rush} et~al.(1993){Rush}, {Malkan} \& {Spinoglio}]{Rush93}
{Rush} B., {Malkan} M.~A., {Spinoglio} L., 1993, \apjs, 89, 1

\bibitem[{Sazonov} et~al.(2007){Sazonov}, {Revnivtsev}, {Krivonos}, {Churazov}
  \& {Sunyaev}]{Sazonov07}
{Sazonov} S., {Revnivtsev} M., {Krivonos} R., {Churazov} E., {Sunyaev} R.,
  2007, \aap, 462, 57

\bibitem[{Simpson}(2005)]{Simpson05}
{Simpson} C., 2005, \mnras, 360, 565

\bibitem[{Stalevski} et~al.(2012){Stalevski}, {Fritz}, {Baes}, {Nakos} \&
  {Popovic}]{Stalevski12}
{Stalevski} M., {Fritz} J., {Baes} M., {Nakos} T., {Popovic} L.~C., 2012,
  Publications de l'Observatoire Astronomique de Beograd, 91, 235

\bibitem[{Suganuma} et~al.(2006){Suganuma}, {Yoshii}, {Kobayashi}
  et~al.]{Suganuma06}
{Suganuma} M., {Yoshii} Y., {Kobayashi} Y., et~al., 2006, \apj, 639, 46

\bibitem[{Tueller}(2011)]{Tueller11}
{Tueller} J., 2011, in { American Astronomical Society Meeting Abstracts
  218\/},  115.02

\bibitem[{Tueller} et~al.(2010){Tueller}, {Baumgartner}, {Markwardt}
  et~al.]{Tueller10}
{Tueller} J., {Baumgartner} W.~H., {Markwardt} C.~B., et~al., 2010, \apjs, 186,
  378

\bibitem[{Turner} et~al.(2009){Turner}, {Miller}, {Kraemer}, {Reeves} \&
  {Pounds}]{Turner09}
{Turner} T.~J., {Miller} L., {Kraemer} S.~B., {Reeves} J.~N., {Pounds} K.~A.,
  2009, \apj, 698, 99

\bibitem[{Ueda} et~al.(2003){Ueda}, {Akiyama}, {Ohta} \& {Miyaji}]{Ueda03}
{Ueda} Y., {Akiyama} M., {Ohta} K., {Miyaji} T., 2003, \apj, 598, 886

\bibitem[{Ueda} et~al.(2007){Ueda}, {Eguchi}, {Terashima} et~al.]{Ueda07}
{Ueda} Y., {Eguchi} S., {Terashima} Y., et~al., 2007, \apjl, 664, L79

\bibitem[{Uttley} et~al.(2003){Uttley}, {Edelson}, {McHardy}, {Peterson} \&
  {Markowitz}]{Uttley03}
{Uttley} P., {Edelson} R., {McHardy} I.~M., {Peterson} B.~M., {Markowitz} A.,
  2003, \apjl, 584, L53

\bibitem[{Vasudevan} et~al.(2010){Vasudevan}, {Fabian}, {Gandhi}, {Winter} \&
  {Mushotzky}]{Vasudevan10}
{Vasudevan} R.~V., {Fabian} A.~C., {Gandhi} P., {Winter} L.~M., {Mushotzky}
  R.~F., 2010, \mnras, 402, 1081

\bibitem[von Ballmoos(2007)]{Von07}
von Ballmoos P., 2007, Focusing Telescopes in Nuclear Astrophysics, Springer

\bibitem[{Wang} \& {Jiang}(2006)]{Wang06}
{Wang} J.~X., {Jiang} P., 2006, \apjl, 646, L103

\bibitem[{Willott} et~al.(2000){Willott}, {Rawlings}, {Blundell} \&
  {Lacy}]{Willott00}
{Willott} C.~J., {Rawlings} S., {Blundell} K.~M., {Lacy} M., 2000, \mnras, 316,
  449

\bibitem[{Winter} et~al.(2009){Winter}, {Mushotzky}, {Reynolds} \&
  {Tueller}]{Winter09}
{Winter} L.~M., {Mushotzky} R.~F., {Reynolds} C.~S., {Tueller} J., 2009, \apj,
  690, 1322

\end{thebibliography}
\end{document}